\documentclass[prl,english,twocolumn,superscriptaddress,showpacs,floatfix,amsfonts]{revtex4}
\usepackage{palatino}
\usepackage{amssymb}
\usepackage{amsmath}
\usepackage{amsthm}
\usepackage{color}
\usepackage{yhmath}
\usepackage{cancel}
\usepackage{graphicx}%
\usepackage{epstopdf}
\makeatletter
\usepackage{babel}
\makeatother
\begin{document}
\title{Validity of  Equation-of-Motion Approach to Kondo Problem in the Large-$N$ limit}
\author{Yunong Qi}
\email{yqi@mail.uh.edu}
\affiliation{Texas Center for Superconductivity, University of Houston, Houston, Texas 77204}
\author{Jian-Xin Zhu}
\email{jxzhu@lanl.gov}
\homepage{http://theory.lanl.gov}
\affiliation{Theoretical Division, Los Alamos National Laboratory, Los Alamos, New Mexico 87545}
\author{C. S. Ting}
\email{csting@mail.uh.edu}
\affiliation{Texas Center for Superconductivity, University of Houston, Houston, Texas 77204}
\date{\today}
\begin{abstract}
The  Anderson impurity model for Kondo problem is investigated for arbitrary orbit-spin degeneracy $N$ of the magnetic impurity by the equation of motion method (EOM).  By employing a new decoupling scheme, a set self-consistent equations for the one-particle Green function are derived and numerically solved in the large-$N$ approximation. For the particle-hole symmetric Anderson model with finite Coulomb interaction $U$, we show that the Kondo resonance at the impurity site exists for all $N \geq 2$. The approach removes the pathology in the standard EOM for $N=2$, and has the same level of applicability as non-crossing approximation.   For $N=2$, an exchange field splits the Kondo resonance into only two peaks, consist with the result from more rigorous numerical renormalization group (NRG) method. The temperature dependence of the Kondo resonance peak is also discussed. 

\end{abstract}
\maketitle
The Kondo effect has been a subject of intensive investigation both experimentally and theoretically for many years. The Anderson impurity model~\cite{Anderson} has been regarded as one of the successful models, which correctly describe the coupling between conduction electrons and local magnetic impurity. Although many techniques have been developed for solving this model~\cite{Hewson}, it is desirable to have a semi-analytic and more generic method that can treat finite $U$ case at finite temperatures. It is believed that the effect due to finite $U$ is important in determining spectroscopic, thermodynamics, and transport properties at finite temperature~\cite{Krishna-murthy,Tsvelick}.

The equations of motion method (EOM)~\cite{Appelbaum1,Lacroix,Czycholl,Qi,Meir91} might be such a candidate. This approach has been employed to derive an analytical expression for the single particle Green's function of the local electron at the impurity site and from which finite temperature properties of the system can be obtained. The EOM successfully yields  approximate but correct behaviors for the resistivity~\cite{Theumann,Appelbaum2}, the spin susceptibility~\cite{Appelbaum2,Mamada}, and other transport properties ~\cite{Poo}.  Especially, when the temperature is above  $T_{K}$, the results based on this approach agree well with those from perturbative calculations. However, the most serious  weakness in the standard EOM is its failure to show the Kondo resonance peak at the Fermi energy for symmetric Anderson model with finite $U$ for spin-orbital $N=2$. Recently, it has been correctly pointed out~\cite{Kashcheyevs} that the particle-hole symmetric case is the singular point for the standard EOM. However, we note that the the non-crossing approximation (NCA) method with large-$N$ expansion ~\cite{Bickers}, when generalized to the finite-$U$ cases~\cite{Pruschke89}, is successful in producing a Kondo resonance peak at the Fermi energy for the particle-hole symmetric case. Therefore, it would be legitimate to ask whether one can develop a large-$N$ EOM and the Kondo resonance peak can be recovered when $N=2$.  Another important issue is whether this kind of method can describe the effect of applied exchange magnetic field correctly. By addressing these two important issues, this Letter will establish the validity of the EOM for treating the Kondo problem in the large $N$ limit.

We start with  the Anderson impurity  Hamiltonian with arbitrary spin-orbital degeneracy $N$ to study the Kondo problem. 
In this model, a single band for conduction electrons is adopted, and the magnetic impurity has $N/2$ degenerate localized orbitals plus each orbital carrying a spin degeneracy of $2$.  
The Anderson Hamiltonian of a magnetic impurity with orbit-spin degeneracy $N$ in metal  has the following expression:
\begin{align}
H =
\displaystyle\sum_{k}\epsilon_{k}c^{\dagger}_{k}c_{k} + \displaystyle\sum^{N}_{\alpha}\epsilon_{\alpha}f^{\dagger}_{\alpha}f_{\alpha} +
U\displaystyle\sum^{N}_{\alpha {\ne} \beta}f^{\dagger}_{\alpha}f_{\alpha}f^{\dagger}_{\beta}f_{\beta} 
\nonumber
\\
+\displaystyle\sum^{N}_{k,\alpha}\left[V_{k\alpha}c^{\dagger}_{k}f_{\alpha} +V_{k\alpha}^{*}
f^{\dagger}_{\alpha}c_{k} \right] \;.
\end{align}
where we have defined $k=(\vec{k},\sigma)$ for conduction band indices.   $c^{\dagger}_{k }$ and $f^{\dagger}_{\alpha}$ are respectively the creation operators for conduction and $f$ electrons at the impurity site. The quantities ${\epsilon}_{k}$, ${\epsilon}_{\alpha}$ are the band energy of  conduction electrons and  energy of the local electron in the $\alpha$-orbital (spin index included) at the impurity site, respectively. For simplicity, a constant conduction electrons density of states is assumed, i.e.,  $\rho(\epsilon)=1/2D$ when $-D {\le} {\epsilon}_{k}{\le} D$, where $D$ is band half-width. Here an $SU(N)$-type  Coulomb interaction with strength $U$ is assumed between electrons of different orbitals (or with opposite spins in the same orbital) at the impurity site, and $V_{k\alpha}$ represents  $s$-$f$ hybridization.   

The decoupling scheme used in  the original work of Appelbaum, Penn, and Lacroix (APL)~\cite{Appelbaum1,Lacroix} has been regarded as standard in EOM.  Using this "standard" scheme for the EOM, one can obtain a coupled set of $N$ self-consistent  equations for the local electron Green's function. In order to make these equations close among themselves in this method, truncation of the perturbative terms has to be made. However, a close look at  the truncating approximation, certain higher order of Green's function terms should not be simply discarded, and they need to be properly included since these terms are absent at $N=2$ and  become important when $N$ is larger than $2$. We here generalize Lacoix's $N=2$  decoupling scheme to large $N$ case and write the matrix element of the following higher order Green's function in the approximate form  
\begin{equation}
{\ll}f_{\alpha}f^{\dagger}_{\gamma}f_{\gamma}c^{\dagger}_{k}f_{\alpha^{\prime}}{\mid}f^{\dagger}_{\beta}{\gg}={\langle}c^{\dagger}_{k}f_{\alpha^{\prime}}{\rangle}{\ll}f_{\alpha}f^{\dagger}_{\gamma}f_{\gamma}{\mid}f^{\dagger}_{\beta}{\gg} + \mathcal{O}(V^{2})
\end{equation}    
Corrections to this decoupling scheme are of order of $V^{2}$ and can be neglected in the limit of $V{\rightarrow}0$ and the validity of this corrections has been discussed by Czycholl~\cite{Czycholl}. Adding the above terms to our work is necessary to make local electron Green's functions to satisfy a set of self-consistent equations and make the EOM more powerful as it will be demonstrated below.  After a complicated and lengthy derivation, the final expression for the matrix element of impurity Green's function $G_{{\alpha}{\beta}}\stackrel{\text{\tiny def}}{=}{\ll}f_{\alpha}{\mid}f^{\dagger}_{\beta}{\gg}{\equiv}G_{\alpha\alpha}\delta_{\alpha\beta}$ in the large-$N$ approximation is found to be
\begin{align}
G_{\alpha\alpha} = 
\dfrac{1 - \bar{n}_{{\alpha}^{\prime}} \left( \omega \right)}{\omega - \epsilon_{\alpha} - {\Sigma}_{0,\alpha} + 
\dfrac{U{\Sigma}_{1,\alpha{\alpha}^{\prime}}}{\omega - \epsilon_{\alpha} - 
U - {\Sigma}_{0,\alpha} - {\Sigma}_{3,\alpha{\alpha}^{\prime}}}} 
\nonumber
\\
+ \dfrac{\bar{n}_{{\alpha}^{\prime}}(\omega)}{\omega - \epsilon_{\alpha} - 
{\Sigma}_{0,\alpha}  - U - \dfrac{U\left({\Sigma}_{3,{\alpha}{\alpha}^{\prime}} - 
{\Sigma}_{1,{\alpha}{\alpha}^{\prime}}\right)}{\omega - \epsilon_{\alpha} - {\Sigma}_{0,\alpha} - {\Sigma}_{3,\alpha{\alpha}^{\prime}}}} 
  \;.  
\label{EQ:dGreen}
\end{align}
Here the spin-orbital index $\alpha^{\prime} {\ne} {\alpha}$ and we have defined several self-energies and functions: 
The  average occupation number of local electrons at the impurity site is defined as
\begin{align}
\bar{n}_{{\alpha}^{\prime}} \left({\omega}\right)  \stackrel{\text{\tiny def}}{=} 
\left(N-1\right)
\biggl{[}
{\langle}n_{{\alpha}^{\prime}}{\rangle} +\displaystyle\sum_{k}\dfrac{V_{k{\alpha}^{\prime}}\langle c^{\dagger}_{k}f_{{\alpha}^{\prime}}\rangle}{D_{1,{\alpha}{\alpha}^{\prime}}(k,\omega)}
\nonumber
\\
+ \displaystyle\sum_{k}\dfrac{V^{*}_{k{\alpha}^{\prime}}\langle f^{\dagger}_{{\alpha}^{\prime}}c_{k}\rangle}{D_{2,{\alpha}{\alpha}^{\prime}}(k,\omega)}
\biggr{]}
 \;,
\end{align}
and the three self-energies are 
\begin{align}
{\Sigma}_{0,\alpha}  \stackrel{\text{\tiny def}}{=} \displaystyle\sum_{k}
\dfrac{\vert V_{k\alpha}\vert^{2}}{{\omega} - {\epsilon}_{k}}\;,
\end{align}

\begin{align}
{\Sigma}_{1,\alpha\alpha^{\prime}} \stackrel{\text{\tiny def}}{=} (N-1)\displaystyle\sum_{k} \dfrac{V^{*}_{k{\alpha}^{\prime}}\biggl{[}
\displaystyle\sum_{q}V_{q\alpha^{\prime}}{\langle} c^{\dagger}_{q}c_{k}{\rangle} - \Sigma_{0,\alpha} {\langle} f^{\dagger}_{\alpha^{\prime}}c_{k}{\rangle} \biggr{]}}{D_{2,{\alpha}{\alpha}^{\prime}}(k,\omega)}
\nonumber
\\
+ (N-1)\displaystyle\sum_{k} \dfrac{V_{k\alpha^{\prime}}\biggl{[} \displaystyle\sum_{q}V^{*}_{q\alpha^{\prime}}{\langle} c^{\dagger}_{k}c_{q}{\rangle} - \Sigma_{0,\alpha}{\langle}c^{\dagger}_{k}f_{\alpha^{\prime}}{\rangle} \biggr{]}}{D_{1,\alpha\alpha^{\prime}}(k,\omega)}
\;,
\label{EQ:Sigma1}
\end{align}
and
\begin{align}
{\Sigma}_{3,\alpha\alpha^{\prime}}  \stackrel{\text{\tiny def}}{=}\displaystyle\sum_{k}\vert V_{k{\alpha}^{\prime}}\vert^{2}\biggl{[}\dfrac{1}{D_{1,{\alpha}{\alpha}^{\prime}}(k,\omega)}+\dfrac{1}{D_{2,{\alpha}{\alpha}^{\prime}}(k,\omega)}\biggr{]}
\;
\label{EQ:Sigma 3} \;,
\end{align}   
where we have defined two functions: $ D_{1,\alpha{\alpha}^{\prime}}\left(k,\omega\right)  \stackrel{\text{\tiny def}}{=} {\omega} +{\epsilon}_{k} - {\epsilon}_{\alpha} - {\epsilon}_{{\alpha}^{\prime}}-U$ 
and $D_{2,\alpha{\alpha}^{\prime}}\left(k,\omega\right)  \stackrel{\text{\tiny def}}{=} {\omega} - {\epsilon}_{k} - {\epsilon}_{\alpha} + {\epsilon}_{{\alpha}^{\prime}}$.
 
The final expression of  Eq.~(\ref{EQ:dGreen}) is nontrivial and requires a delicate derivation. We can simply summarize our main results: 1) High order Green's functions have been naturally included in our derivation, which enables us to truncate arbitrary order to close the coupling integral equations;  2) A higher order decoupling scheme we have used in our derivation makes our formula distinguish from the previous conventional $N=2$ results; 3) Eq.~(\ref{EQ:dGreen}) is not an extension to APL~\cite{Appelbaum1,Lacroix}, but a new large-$N$ decoupling scheme. It is due to the fact all of the terms $N-1$, which are implicitly included in Eq.~(\ref{EQ:dGreen}) and explicitly expressed in Eqs. (4) and (6) , are absent if one takes APL's decoupling with large-$N$ EOM.  

We shall illustrate the new features and ingredients of Eq.~(\ref{EQ:dGreen}): 1) It includes a set of the $N{\ge}2$ closed self-consistent integral equations,which can be numerically solved; 2) The effective occupancy is frequency dependent, 3) The higher order self-energy contains the intermediate off-diagonal states in momentum space (e.g., $\langle c_{k}^{\dagger}c_{q} \rangle$) and charge fluctuations (e.g., $\langle f_{\alpha}^{\dagger}c_{q} \rangle$); 4) It reproduces the results from the standard $N=2$ EOM in the infinite-$U$ limit~\cite{Qi}.

Before we proceed to carry out numerical calculations, we shall also point out that the expectation value of ${\langle}c^{\dagger}_{q}c_{k}{\rangle}$ and  ${\langle}f^{\dagger}_{{\alpha}^{\prime}}c_{k}{\rangle}$, which have been discarded in the EOM with APL decoupling scheme prove to be very important at low temperatures since they diverge logarithmically at the Fermi level as the temperature approaches to zero. Their values should be self-consistently evaluated through the spectral densities between the conduction electron and the impurity Green's functions.
 \begin{equation}
 {\langle}c^{\dagger}_{q}c_{k}{\rangle} =
- \dfrac{1}{\pi}{\displaystyle\int}f_{FD}(\omega)\mbox{Im}{\ll}c_{k}{\mid}c^{\dagger}_{q}{\gg}d{\omega}\;,
\end{equation}      
where $f_{FD}(\omega)= 1/[\exp(\omega/k_{B}T) + 1]$ is the Fermi-Dirac distribution function, and
the Green's function ${\ll}c_{k\bar{\sigma}}{\mid}c^{\dagger}_{q\bar{\sigma}}{\gg}$ is:
\begin{equation}
{\ll}c_{k}{\mid}c^{\dagger}_{q}{\gg} =\dfrac{{\delta}_{q,k}}{\omega - {\epsilon}_{k}} +
\dfrac{V_{k{\alpha}^{\prime}}V_{q{\alpha}^{\prime}}^{*}{\ll}f_{{\alpha}^{\prime}}{\mid}f^{\dagger}_{{\alpha}^{\prime}}{\gg}}{\left(
\omega - {\epsilon}_{k}\right)\left(\omega - {\epsilon}_{q}\right)}\;,
\end{equation}
and similarly
\begin{equation}
{\langle}f^{\dagger}_{{\alpha}^{\prime}}c_{k}{\rangle} =
-\dfrac{1}{\pi}{\displaystyle\int}f(\omega)\mbox{Im}{\ll}c_{k}{\mid}f^{\dagger}_{{\alpha}^{\prime}}{\gg}d{\omega} \;,
\end{equation}
with 
\begin{equation}
{\ll}c_{k}{\mid}f^{\dagger}_{{\alpha}^{\prime}}{\gg} =  
\dfrac{V_{k{\alpha}^{\prime}}{\ll}f_{{\alpha}^{\prime}}{\mid}f^{\dagger}_{{\alpha}^{\prime}}{\gg}}{\left(
\omega - {\epsilon}_{k}\right)}\;.
\end{equation}
Solving the coupled Eq.~(\ref{EQ:dGreen}) not only yields the correct Kondo resonance at low temperatures but also allows us to explicitly include the logarithmic divergence in general. When the spin-dependent effect is taken into account, the significance of our new EOM approach is that it does not rely on the additional renormalization introduced in the previous EOM technique~\cite{Meir91}. The purpose of the additional renormalization is to account for the spin dependent level splitting and broadening ~\cite{Meir93,Martinek1}. The lack of rigorous justification for the existence of the additional renormalization has cast a doubt for the effectiveness of the EOM approach for the non-equilibrium Kondo problem. In our improved EOM formula, we find that the correct Kondo resonance can be derived without introducing the additional renormalization. Comparing with previous calculations, we have properly evaluated terms such as ${\langle}c^{\dagger}_{q}c_{k}{\rangle}$, ${\langle}f^{\dagger}_{{\alpha}^{\prime}}c_{k}{\rangle}$ through Eq.~(\ref{EQ:Sigma1})~\cite{Qi}. These terms make crucial contributions to the Kondo resonance peak at very low temperatures. Neglecting these terms will lead to severe errors, which has to be recovered by  adding {\em ad hoc} an additional renormalization. 
\begin{figure}[t]
\begin{center}
\includegraphics[width=8cm]{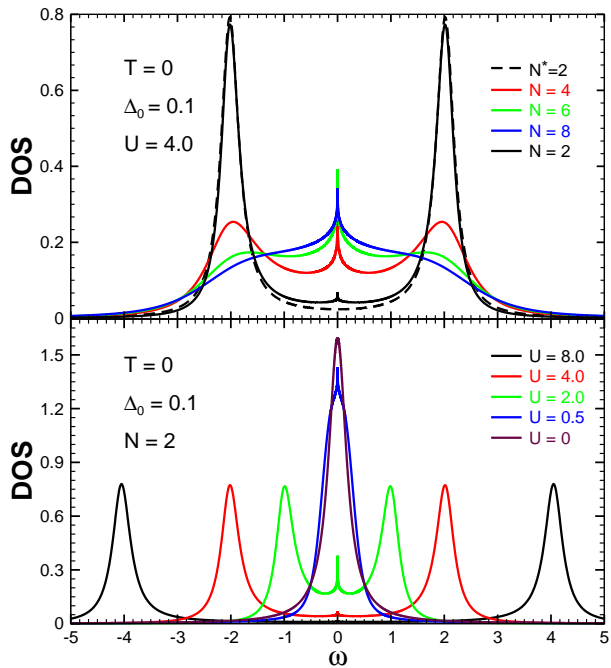}
\caption{(Color) (top): Spectral density DOS calculated via the EOM method for particle-hole symmetric Anderson impurity model for different orbit-spin degeneracy $N$ at zero temperature with Coulomb interaction $U=4$, solid lines correspond to our large $N \ge 2$ formulation and dash line is the previous $N=2$ formulation result; (bottom) Results of spectral density DOS via large $N$ limit EOM for $N=2$ particle-hole symmetric Anderson model with different Coulomb interaction $U$}
\label{fig:fig1.eps}
\end{center}
\end{figure}

In order to demonstrate the power of our large-$N$ EOM approach, we shall apply our new formulation to consider the Kondo impurity problem in the particle-hole symmetric case, where  the standard $N=2$ EOM fails to show the Kondo resonance peaks.  In this case,  the impurity levels for $V_{k\alpha}=0$ are symmetrically placed about the Fermi level at ${\epsilon}=\epsilon_{F}-U/2$ and ${\epsilon}=\epsilon_{F}+U/2$. If the conduction bands are symmetric about the Fermi level and half-filled, the model has complete particle-hole symmetry and the average occupation of each spin-orbital ${\langle}n_{\alpha}{\rangle}=1$. This symmetric model can display the full range of behavior from non-magnetic for $k_{B}T, U <{\Delta}_{0}$, to magnetic and Kondo behavior for $U {\gg} {\Delta_{0}}$, where $\Delta_{0}= - \mbox{Im} [\Sigma_{0,\alpha}(\omega+i0^{+})]$. We shall examine this well-studied case by numerically solving Eq.~(\ref{EQ:dGreen}) and we choose the following parameters for our numerical calculation: The energy of the half-width of the impurity resonance in a nonmagnetic metal,  $\Delta_{0}$ is taken $0.1$ in the unit of conduction band half-width $D$ unless specified otherwise. In this special case,  the Coulomb interaction energy $U$ has usually been taken as a parameter. 

The first illuminating example is the more familiar equilibrium Kondo problem where neither impurity state nor hybridization is spin-dependence and spin-orbital degeneracy $N$. As shown on the top panel of Figure~\ref{fig:fig1.eps}, when one uses the formulation from the standard EOM for $N=2$~\cite{Qi},  there is no Kondo resonance peaks (dashed line).  This result numerically confirms the analysis made 
by Kashcheyevs {\em et al.}~\cite{Kashcheyevs} that the standard EOM technique, as a severe drawback,  cannot produce the Kondo resonance at the particle-hole symmetric point. 
Interestingly, within our large-$N$ EOM technique, the Kondo resonance peak at the Fermi energy is indeed obtained even for $N=2$.  With increase of  the spin-orbital degeneracy $N$, the spectral weight is transferred more significantly from the virtual bound states toward the coherence region around the Fermi energy. This behavior is similar to the case, as demonstrated below, when the localized level is moved toward the Fermi energy within the energy region of bare resonance width $\Delta_{0}$ --- mixed valence region. 
The bottom panel of the Fig.~\ref{fig:fig1.eps} shows the impurity spectral density ($\rho_{f\uparrow}=\rho_{f\downarrow}$) DOS for different Coulomb energy $U$ at zero temperature by using our large-$N$ EOM  formula. At small values of $U$, the DOS exhibits  the two broad peaks at ${\epsilon}_{\alpha}=-U/2$ and ${\epsilon}_{\alpha}=U/2$, respectively,  and a sharp Kondo resonance peak at the Fermi level.
With the increased $U$, the Kondo resonance peak disappears gradually, evolving into a localized state. These well-known results agree with many various approaches, e.g., the scaling analysis~\cite{Haldane}, the NRG method and NCA~\cite{Bickers}.

\begin{figure}[t]
\begin{center}
\includegraphics[width=8cm]{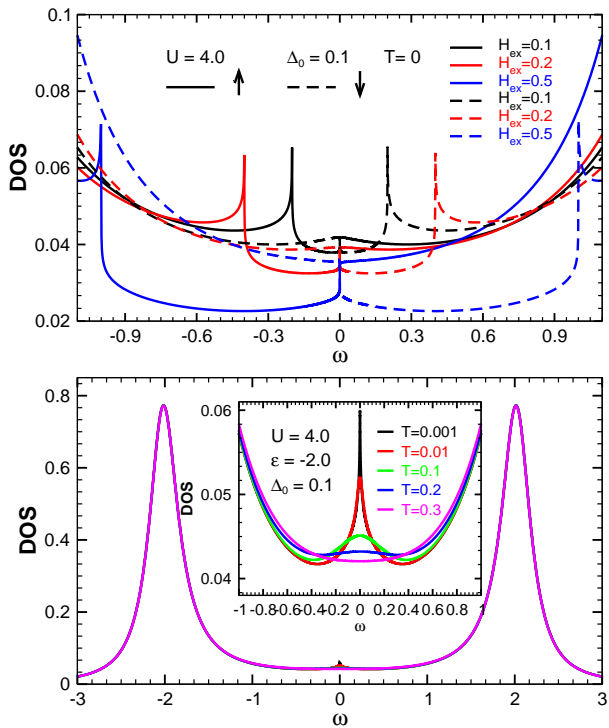}
\caption{(Color) Spectral density DOS calculated via the large $N$ limit EOM method $N=2$ for a finite $U$ patrticle-hole symmetric Anderson impurity model at different exchange fields (top) as well as  at various temperatures(bottom). The inset displays the zoom-in view of the Kondo resonance near the Fermi energy.}
\label{fig:fig2.eps}
\end{center}
\end{figure}
We now consider another very interesting application of our large-$N$ limit EOM technique to the Kondo impurity problem in the presence of external magnetic field.  
We show, in the top panel of Fig.~\ref{fig:fig2.eps},  the spectral density DOS for a finite $U$ particle-hole symmetric model at different exchange fields. A splitting of the Kondo resonances peaks for the spin-up and down electrons is obtained at energy ${\epsilon}={\pm}2H_{ex}$.  There is no spurious peak but only a small bump showing up near the Fermi energy. This result is in reasonable agreement with that from the NRG calculations~\cite{Costi00,Hewson06}, except for a small bump obtained here. The temperature dependent effect has been shown in the bottom panel of Fig.~\ref{fig:fig2.eps} . We find that the width and height of Kondo resonance peaks dramatically changed with increasing temperature. The inset displays the zoom-in view of the Kondo resonance near the Fermi energy. By further increasing the temperature, the Kondo resonance peaks disappear at a characteristic temperature (about 0.3 for the given parameter values).  We should also mention but not show that the new large-$N$ EOM technique can describe equally well the Kondo physics for an asymmetric Anderson impurity model.

In conclusion, we have developed a large-$N$ EOM approach to the Kondo impurity problem for arbitrary Coulomb interaction $U$ at finite temperatures. Numerical results  are carried out  for symmetric Anderson impurity model with finite $U$. We show that  the Kondo resonance peak, which escapes from the standard EOM approach for the particle-hole symmetric point, can be restored in the new technique.   Furthermore, we have also shown that the new technique describes reasonably well the field dependence of the Kondo effect.  Both successes establish the power of this new EOM technique.   

We here should give special thanks to Shufeng Zhang and R. C Albers for useful discussions. This work was supported by the Robert Welch Foundation No. E-1146 at the University of Houston (Y.Q. and C.S.T.), by U.S. DOE under Contract No. DE-AC52-06NA25396 and under Grant Nos.  LDRD-DR X9GT \& X9HH  (J.X.Z.).

\bibliography{}

\end{document}